\documentclass[manuscript,screen,nonacm]{acmart}
\AtBeginDocument{%
  \providecommand\BibTeX{{%
    \normalfont B\kern-0.5em{\scshape i\kern-0.25em b}\kern-0.8em\TeX}}}

\begin{document}

\title{Designing and Evaluating Scalable Privacy Awareness and Control User Interfaces for Mixed Reality}
%
\author{Marvin Strauss}
\email{marvin.strauss@uni-due.de}
\orcid{0009-0007-1040-9175}
\affiliation{%
    \institution{Human-Computer Interaction Group, University of Duisburg-Essen}
    \city{Essen}
    \country{Germany}
    }

\author{Viktorija Paneva}
\email{viktorija.paneva@unibw.de}
\orcid{0000-0002-5152-3077}
\affiliation{%
    \institution{Usable Security and Privacy Group, University of the Bundeswehr Munich}
    \city{Munich}
    \country{Germany}
    }

\author{Florian Alt}
\email{florian.alt@unibw.de}
\orcid{0000-0001-8354-2195}
\affiliation{%
    \institution{Usable Security and Privacy Group, University of the Bundeswehr Munich}
    \city{Munich}
    \country{Germany}
    }

\author{Stefan Schneegass}
\orcid{0000-0002-0132-4934}
\email{stefan.schneegass@uni-due.de}
\affiliation{%
    \institution{Human-Computer Interaction Group, University of Duisburg-Essen}
    \city{Essen}
    \country{Germany}
    }

\renewcommand{\shortauthors}{Strauss}

\begin{abstract}
As Mixed Reality (MR) devices become increasingly popular across industries, they raise significant privacy and ethical concerns due to their capacity to collect extensive data on users and their environments. This paper highlights the urgent need for privacy-aware user interfaces that educate and empower both users and bystanders, enabling them to understand, control, and manage data collection and sharing. Key research questions include improving user awareness of privacy implications, developing usable privacy controls, and evaluating the effectiveness of these measures in real-world settings. The proposed research roadmap aims to embed privacy considerations into the design and development of MR technologies, promoting responsible innovation that safeguards user privacy while preserving the functionality and appeal of these emerging technologies.
\end{abstract}

\maketitle

\section{Motivation}

In the realm of Extended Reality (XR), Mixed Reality (MR) headsets represent a technological convergence, bridging Augmented Reality (AR) and Virtual Reality (VR)~\cite{rauschnabel2022,speicher_what_2019}. 
These headsets offer an unprecedented level of immersion and interactivity, catering to a wide array of applications across various industries~\cite{park2022metaverse}. 
The versatility of MR headsets, ranging from entertainment to professional training, positions them as a pivotal component in the next wave of digital transformation. 
However, the rapid evolution of these technologies necessitates a proactive approach to address the emerging challenges they present, particularly in the domains of privacy, ethics, and user safety~\cite{bye2019ethical}.

The escalating adoption of MR headsets brings to the fore significant privacy concerns~\cite{guzman2023privacy}. 
These devices, by their very nature, are capable of capturing extensive data about users, their environment, and their surroundings~\cite {nair_exploring_2023}. 
This data, if misused, could lead to unsolicited profiling or surveillance, posing a threat to individual privacy and societal norms~\cite{tricomi_you_2023}. 
The potential for such misuse underscores the urgent need for policies that safeguard against these risks~\cite{adams_2018}.

Key challenges in this domain include raising users’ and bystanders’ awareness and increasing their understanding of the privacy implications of XR technologies \cite{rauschnabel_antecedents_2018, ohagan_privacy-enhancing_2023}. 
Moreover, the investigation of their privacy perception towards these devices is important to understand their concerns and desires. 
Consequently, it is crucial to protect all parties from the unaware or unwanted collection of sensitive data. 
This protection must be balanced with the continued innovation and utility of XR technologies.

\section{Open Research Questions}
In the context of the challenges associated with increasing awareness of data collection and privacy implications on MR devices, the following open research questions emerge:

\begin{enumerate}
  \item How can MR user interfaces raise awareness among users about the data being collected, processed, and shared?
  \item How can users’ MR privacy behavior be investigated in realistic settings?
  \item How can bystanders of MR users be informed of ongoing tracking and be granted control over their data?
  \item How can MR user interfaces efficiently communicate privacy risks associated with consenting to data collection and sharing at opportune moments?
  \item How can MR user interfaces support efficient privacy permission control, including understanding, granting, reviewing, and revoking permissions?
  \item How can the impact of MR privacy interfaces be assessed?
  \item How can researchers and practitioners be supported in the privacy-preserving design of MR applications?
\end{enumerate}

\section{Research Roadmap}

We sketch a research roadmap to address the aforementioned research questions.
Firstly, a profound \textbf{understanding of users’ awareness, mental models, and their understanding of the implications of using MR technology on their privacy} needs to be obtained. This knowledge is crucial for informing the design of privacy-preserving user interfaces in a way that raises users’ awareness, supports users in building up the required proficiency, and gives users control and feedback mechanisms to make strong and informed decisions.

Secondly, \textbf{usable privacy control UIs for MR applications and devices} need to be created. 
As many privacy mechanisms are deliberately designed with low usability, that is, designers increase the interaction costs in terms of time and effort in a way that users ignore them or deliberately violate UI guidelines to bias users’ decisions (cf. cookie banners, privacy policies, and permission systems). 
As a result, facilitating the design of user interfaces that (a) minimize the effort for users, and (b) enable strong and confident privacy decisions, scaling to the ever-increasing number of devices and applications becoming available for MR, is key.

Thirdly, the developed \textbf{privacy control UIs need to be evaluated}. 
On one hand, there is a need to rigorously measure the usability of the concepts, quantifying how easy they are to learn, how efficiently they can be used, how easy they make it for users to memorize privacy decisions, and how satisfied users are. 
On the other hand, an interesting question is how concepts affect privacy behavior for other technologies, how users’ self-efficacy evolves, and how interfaces can be designed not to make users dependent.

Finally, a core challenge in evaluating privacy user interfaces regarding long-term effects is creating an environment in which users behave naturally and in an unbiased way. 
A \textbf{real-world testbed} is required where privacy user interfaces can be evaluated in users’ everyday lives as they interact with MR technology.

By fulfilling these objectives, it becomes possible to embed privacy as an integral aspect of the design of MR applications and provide a valuable resource for researchers, practitioners, and manufacturers that empowers them to address privacy challenges during the design and development phases rather than as an afterthought.

\section{Conclusion}
To conclude, the widespread adoption of XR technologies, particularly MR, presents opportunities and challenges. 
As we stand at the cusp of this technological revolution, it is crucial to develop policy frameworks that not only encourage responsible innovation but also address potential vulnerabilities. 
With our research, we aim to take a proactive step in this direction, aiming to integrate privacy considerations into the fabric of MR and XR technology design and development. 
By participating in this workshop, we aim to contribute to the collaborative effort of charting a responsible and sustainable course for the XR landscape, ensuring that innovation does not come at the cost of privacy and ethical considerations.

\bibliographystyle{ACM-Reference-Format}
\bibliography{sample-base}

\end{document}